\documentclass[12pt,english]{paper}
\usepackage{amsthm,amsmath,amssymb}
\usepackage[T1]{fontenc}
\usepackage[latin1]{inputenc}
\usepackage{graphicx}
\usepackage{cite}
\makeatletter

\newcommand{\bibstyle@aas}{\bibpunct{(}{)}{;}{a}{}{,}}
\makeatletter

\makeatletter

\usepackage{geometry}

\makeatletter

\usepackage{epsfig}

\makeatother

\makeatother

\makeatother

\usepackage{babel}
\makeatother
\begin{document}

\title{Measurement of high energy dark matter from the Sun at IceCube}

\author{Ye Xu$^{1,2}$}

\maketitle

\begin{flushleft}
$^1$School of Electronic, Electrical Engineering and Physics,  Fujian University of Technology, Fuzhou 350118, China
\par
$^2$Research center for Microelectronics Technology, Fujian University of Technology, Fuzhou 350118, China
\par
e-mail address: xuy@fjut.edu.cn
\end{flushleft}

\begin{abstract}
It is assumed that heavy dark matter particles (HDMs) with a mass of O(TeV) are captured by the Sun. HDMs can decay to relativistic light dark matter particles (LDMs), which could be measured by km$^3$ neutrino telescopes (like the IceCube detector). The numbers and fluxes of expected LDMs and neutrinos were evaluated at IceCube with the $Z^{\prime}$ portal dark matter model. Based on the assumption that no events are observed at IceCube in 6 years, the corresponding upper limits on LDM fluxes were calculated at 90\% C. L.. These results indicated that LDMs could be directly detected in the O(1TeV)-O(10TeV) energy range at IceCube with 100 GeV $\lesssim m_{Z^{\prime}} \lesssim$ 350 GeV and $\tau_{\phi} \lesssim 5\times10^{22}$ s.
\end{abstract}

\begin{keywords}
High energy dark matter, Dark matter accumulation, Neutrino
\end{keywords}

\section{Introduction}
It was found in cosmological and astrophysical observations that the bulk of matter in the Universe consists of dark matter (DM). $84\%$ of the matter content is thermal DM in the
Universe, which were created thermally in Early Universe\cite{bergstrom,BHS,Planck2015}. The searches for high energy (from O(GeV) to O(TeV)) neutrinos which are produced by the DM annihilation in the Sun's core have been performed using the data recorded by the IceCube and ANTARES neutrino telescopes\cite{icecubedm-sun,antaresdm-sun}. But no one has found thermal DM particles yet\cite{XENON1T,PANDAX,fermi,antares-icecube-dm-MW,icecubedm-sun,antaresdm-sun}.
\par
The heavy dark sector with a mass of O(TeV) is an alternative DM scenario\cite{KC87,CKR98,CKR99,KT,KCR,CKRT,CCKR,KST,CGIT,FKMY,FKM}. In this model, there exist at least two DM species in the Universe (for example, heavy and light DM particles). Heavy dark matter (HDM), $\phi$, is a thermal particle which is generated by the early universe. The bulk of present-day DM consists of them. The other is a stable light dark matter particle (LDM), $\chi$ which is the product of the decay of HDM ($\phi\to\chi\bar{\chi}$). Due to the decay of long-living HDMs ($\tau_{\phi} \gg t_0$\cite{AMO,EIP}, $t_0$ is the age of the Universe), the present-day DM may also contain a small component which is high energy LDMs. Besides direct measurements of HDMs, one can detect the standard model (SM) products of decay of HDMs. The search for these high energy SM particles from the Sun (they should be neutrinos in this measurement) has been performed using the data recorded by the IceCube neutrino observatory\cite{icecubedm-decay}. In this work, however, the products of the decay of HDMs are a class of LDMs\cite{BGG,BGGM}, not SM particles.
\par
The LDMs from the Sun's core could be more easily detected with IceCube, compared to those from the Earth's core, since the HDM accumulation in the Sun is much greater than that in the Earth\cite{BCH,LE}. LDMs would interact with nuclei when they pass through the Sun, the Earth and ice. %A Z$^{\prime}$ portal dark matter model\cite{APQ,Hooper} will be taken for LDMs $\chi$ to interact with nuclei in this work.
Those LDMs can be directly measured with the IceCube neutrino telescope. The capability of the measurement of those LDMs will be discussed here. In this measurement, the background consists of muons and neutrinos generated in cosmic ray interactions in the Earth's atmosphere and astrophysical neutrinos.
\par
In what follows, the distributions and numbers of expected LDMs and neutrinos will be evaluated in the energy range 1-100 TeV assuming 6 years of IceCube data. Then the upper limits on LDM fluxes were also calculated at 90\% C.L.. Finally, the capability of the measurement of TeV LDMs will be evaluated at IceCube.
\par
\section{HDM accumulation in the Sun}
HDMs of the Galactic halo would collide with atomic nuclei in the Sun when their wind sweeps through the Sun. A fraction of those HDMs would lose enough kinetic energy to be trapped in orbit. With further collisions with atomic nuclei in the Sun's interior, they would eventually thermalize and
settle in the Sun's core under the influence of gravitation of the Sun's interior. Those HDMs inside the Sun can decay into LDMs at an appreciable rate. Then the number N of HDMs, captured by the Sun, is obtained by the following equation\cite{BCH}
\par
\begin{center}
\begin{equation}
\frac{dN}{dt}=C_{cap}-2\Gamma_{ann}-C_{evp}N-C_{dec}N
\end{equation}
\end{center}
\par
where $C_{cap}$, $\Gamma_{ann}$ and $C_{evp}$ are the capture rate, the annihilation rate and the evaporation rate, respectively. The evaporation rate is only relevant when the DM mass < 5 GeV\cite{BCH}, which are much lower than my interested mass scale (the mass of HDM, m$_{\phi}$ $\geq$ 1 TeV). Thus their evaporation contributes to the accumulation in the Sun at a negligible level in the present work. $C_{dec}$ is the decay rate for HDMs. Since the fraction of HDM decay $\leq$ 3.2$\times 10^{-14}$ per year ($\tau_{\phi} \geq 10^{21}$ s), its contribution to the HDM accumulation in the Sun can be ignored in the evaluation of HDM accumulation. $\Gamma_{ann}$ is obtained by the following equation\cite{BCH}
\par
\begin{center}
\begin{equation}
\Gamma_{ann}=\frac{C_{cap}}{2}tanh^2\left(\frac{t}{\tau}\right)\approx \frac{C_{cap}}{2} \quad with \quad t\gg\tau
\end{equation}
\end{center}
\par
where $\tau=(C_{cap}C_{ann})^{-\frac{1}{2}}$ is a time-scale set by the competing processes of capture and annihilation. At late times $t\gg\tau$ one can approximate tanh$^2\displaystyle\frac{t}{\tau}$=1 in the case of the Sun\cite{BCH}.
%Taking $t$ (about 10$^{17}$s) the age of the Earth, $\displaystyle\frac{t}{\tau}$ can be obtained by the following function\cite{LLL}
%\par
%\begin{center}
%\begin{equation}
%\frac{t}{\tau}\approx1.9\times10^4\left(\frac{C_{cap}}{s^{-1}}\right)^{1/2}\left(\frac{\left\langle\sigma_{ann}\upsilon\right\rangle}{cm^3s^{-1}}\right)^{1/2}\left(\frac{m_{HDM}}{10 %GeV}\right)^{3/4}
%\end{equation}
%\end{center}
%%\par
%where $\left\langle\sigma\upsilon\right\rangle \sim 3\times10^{-26}cm^3s^{-1}$ for DM that are a thermal relic\cite{BHS}.
$C_{cap}$ is proportional to $\displaystyle\frac{\sigma_{\phi N}}{m_{\phi}}$\cite{BCH,JKK}, where $m_{\phi}$ is the mass of HDM and  $\sigma_{\phi N}$ is the scattering cross section between the nucleons and HDMs. The spin-independent cross section is only considered in the capture rate calculation. Then $\sigma_{\phi N}$ is taken to be 10$^{-44}$ cm$^2$ for $m_{\phi} \sim$ O(TeV) \cite{XENON1T,PANDAX}.
\par
The HDM distribution in the Sun is obtained by\cite{BCH},
\par
\begin{center}
\begin{equation}
n(r)=n_0exp\left(-\frac{r^2}{r_{\phi}^2}\right),  \quad with\quad r_{\phi}=\left(\frac{3T_s}{2\pi G_N\rho_s m_{\phi}}\right)^{1/2}\approx0.01R_{sun}\sqrt{\frac{100TeV}{m_{\phi}}}
\end{equation}
\end{center}
\par
where G$_N$ is the Newtonian gravitational constant. $\rho_s\approx$ 151 g/cm$^3$ and $T_s\approx$ 15.5$\times 10^6$ K are the matter density and temperature at the sun center, respectively. R$_{sun}$ is the radius of the Sun. One finds that HDMs are concentrated around the center of the Sun.

\section{LDM and neutrino interactions with nuclei}
In this work, a $Z^{\prime}$ portal dark matter model\cite{APQ,Hooper} is taken for LDMs to interact with nuclei via a neutral current interaction mediated by a gauge boson $Z^{\prime}$ which couples to both the LDMs and quarks (see Fig. 1a in Ref.\cite{BGG}). Here a LDM is assumed to be a Dirac fermion. As assumed in Ref.\cite{BGG}, the interaction vertexes ($\chi\chi Z^{\prime}$ and $qqZ^{\prime}$) are vector-like in this model, since $Z^{\prime}$ vector boson typically acquires mass through the breaking of an additional U(1) gauge group at high energies. This deep inelastic scattering (DIS) cross-section for $\chi+N \to \chi+anything$ ($N$ is a nucleus) is computed in the same way as the Ref.\cite{BGG} in this work. The effective interaction Lagrangian can be written as follows:
\begin{center}
\begin{equation}
\mathcal{L} = \bar{\chi}g_{\chi\chi Z^{\prime}}\gamma^{\mu}\chi Z^{\prime}_{\mu} + \sum_{q_i} \bar{q_i}g_{qqZ^{\prime}}\gamma^{\mu}q_iZ^{\prime}_{\mu}
\end{equation}
\end{center}
where $q_i$'s are the SM quarks, and $g_{\chi\chi Z^{\prime}}$ and $g_{qqZ^{\prime}}$ are the $Z^{\prime}$-$\chi$ and $Z^{\prime}$-$q_i$ couplings, respectively. This Deep inelastic scattering (DIS) cross-section is computed in the lab-frame using tree-level CT10 parton distribution functions\cite{LGH}. The coupling constant G ($G=g_{\chi\chi Z^{\prime}}g_{qqZ^{\prime}}$) is chosen to be 0.05. The masses of $Z^{\prime}$ are taken to be 100 GeV, and 250 GeV and 350 GeV, respectively. Here the mass of LDM $m_{\chi}$ is assumed to be 8 GeV, then the outgoing energy of LDM caused by the decay of HDM  E$_{\chi}\approx \displaystyle\frac{1}{2}m_{\phi}$. The computed DIS cross section obeys a simple power-law form for the energies between 1 TeV and 1PeV.
%The LDM-nucleus cross section $\propto$ $\displaystyle\frac{G^2\mu_{\chi N}^2\nu^2}{\pi m_{Z^{\prime}}^4m_{\chi}^2}$, where $\mu_{\chi N}$ is the reduced mass of the LDM-nucleus system and $\nu$ is the velocity of LDM in the lab frame(see Ref.\cite{APQ,Hooper}).
With m$_{Z^{\prime}}$ = 250 GeV, for example, its cross section is obtained by the following function:
\begin{center}
\begin{equation}
\sigma_{\chi N}=1.284\times10^{-41} cm^2 \left(\frac{E_{\chi}}{1GeV}\right)^{0.970}
\end{equation}
\end{center}
where E$_{\chi}$ is the LDM energy.
\par
The DIS cross-section for neutrino interaction with nuclei is computed in the lab-frame and given by simple power-law forms\cite{BHM} for neutrino energies above 1 TeV:
\begin{center}
\begin{equation}
\sigma_{\nu N}(CC)=4.74\times10^{-35} cm^2 \left(\frac{E_{\nu}}{1 GeV}\right)^{0.251}
\end{equation}
\end{center}
\par
\begin{center}
\begin{equation}
\sigma_{\nu N}(NC)=1.80\times10^{-35} cm^2 \left(\frac{E_{\nu}}{1 GeV}\right)^{0.256}
\end{equation}
\end{center}
where $\sigma_{\nu N}(CC)$ and $\sigma_{\nu N}(NC)$ are the DIS cross-sections for neutrino interaction with nuclei via a charge current (CC) and neutral current (NC), respectively. $E_{\nu}$ is the neutrino energy.
\par
The inelasticity parameter $y=1 - \displaystyle\frac{E_{\chi^{\prime},lepton}}{E_{in}}$ (where $E_{in}$ is the incoming LDM or neutrino energy and $E_{\chi^{\prime},lepton}$ is the outgoing DM particles or lepton energy). $E_{sec}=yE_{in}$, where $E_{sec}$ is the secondaries' energy after a LDM or neutrino interaction with nuclei. The mean values of $y$ for LDMs have been computed:
\begin{center}
\begin{equation}
\left\langle y \right\rangle =\frac{1}{\sigma(E_{in})} \int^1_0 y \frac{d\sigma}{dy}(E_{in},y)dy
\end{equation}
\end{center}
%The mean values of $y$ for LDMs with energies from 1 TeV to 200 TeV are near 0.43 and the ones for neutrinos (NC) vary from 0.46 to 0.34. These results are similar to the results in the Refs.\cite{BGG,GQRS}
The LDM and neutrino interaction lengths can be obtained by
\begin{center}
\begin{equation}
L_{\nu,\chi}=\frac{1}{N_A\rho\sigma_{\nu,\chi N}}
\end{equation}
\end{center}
where $N_A$ is the Avogadro constant, and $\rho$ is the density of matter, which LDMs and neutrinos interact with.
\section{Flux of LDMs which reach the Earth}
The LDMs which reach the Earth are produced by the decay of HDMs in the Sun's core. These LDMs have to pass through the Sun and interact with nuclei inside the Sun. Then the number N$_s$ of LDMs which reach the Sun's surface is obtained by the following equation:
\begin{center}
\begin{equation}
\begin{aligned}
N_s &=2N_0\left(exp(-\frac{t_0}{\tau_{\phi}})-exp(-\frac{t_0+T}{\tau_{\phi}})\right)\prod_{i=1}^{n=\mathcal{N}} exp(\frac{\delta L}{L_i}) \qquad with \quad T \ll \tau_{\phi}\\
    &\approx 2N_0\frac{T}{\tau_{\phi}}exp(-\frac{t_0}{\tau_{\phi}})\prod_{i=1}^{n=\mathcal{N}} exp(\frac{\delta L}{L_i})
\end{aligned}
\end{equation}
\end{center}
where N$_0$=$\displaystyle\int^{t_s}_0 \displaystyle\frac{dN}{dt} dt$ is the number of HDMs captured in the Sun. t$_s$ and t$_0$ are the ages of the Sun and the Universe, respectively. T is the lifetime of taking data for IceCube and taken to be 6 years. If the distance from the Sun's center to the Sun's surface is equally divided into $\mathcal{N}$ portions, $\delta L=\displaystyle\frac{R_{sun}}{\mathcal{N}}$. $L_i=\displaystyle\frac{1}{N_A\rho_i\sigma_{\chi N}}$ is the LDM interaction length at i$\times\delta L$ away from the Sun's center. $\rho_i$ is the density at i$\times\delta L$ away from the Sun's center\cite{SN}. N$_s$ is computed in column density in the present work. The first exponential term  %$\left(exp(-\displaystyle\frac{t_0}{\tau_{\phi}})-exp(-\displaystyle\frac{t_0+T}{\tau_{\phi}})\right)$
in the Eqn. 10 is the fraction of decay of HDMs in the Sun's core. The term of continued product %$\displaystyle\prod_{i=1}^{n=\mathcal{N}} exp(\displaystyle\frac{\delta L}{L_i})$
in the Eqn. 10 is the faction of LDMs which reach the Sun's surface. Here $\mathcal{N}$ is taken to be 10$^4$. The results with $\mathcal{N}$=10$^4$ is sufficiently accurate, whose uncertainty is about 0.05\%.
\par
Then the flux $\Phi_{LDM}$ of LDMs, which reach the Earth, from the Sun's core is described by
\begin{center}
\begin{equation}
\Phi_{LDM}=\frac{N_s}{4\pi D_{se}^2}
\end{equation}
\end{center}
where $D_{se}$ is the distance between the Sun and Earth.
\section{Evaluation of the numbers of expected LDMs and neutrinos at IceCube}
The lifetime for HDMs decaying into SM particles is strongly constrained ($\tau \geq$ O($10^{26}-10^{29}$)s) by diffuse gamma and neutrino observations\cite{EIP,MB,RKP,KKK}. Since the present work considers an assumption that HDMs are unable to decay to SM particles, the constraints on the lifetime for HDM are only those based on cosmology (the age of the Universe is about $10^{17}$ s). Since $\tau_{\phi} \gg 10^{17}$ s in the $Z^{\prime}$ portal dark matter model\cite{AMO,EIP}, $\tau_{\phi} \geq 10^{21}$ s in this work.
\par
IceCube is a km$^3$ neutrino telescope and deployed in the deep ice below the geographic South Pole\cite{icecube2004}. It can detect neutrino interactions with nuclei via the measurement of the cascades caused by their secondary particles above the energy threshold of 1 TeV\cite{icecube2019}. The LDMs from the Sun, which pass through the IceCube detector, would interact with the nuclei inside IceCube. This is very similar to the DIS of neutrino interaction with nuclei via a neutral current, whose secondary particles would develop into a cascade at IceCube.
\par
In this analysis, LDM events were selected with the following event selection criteria. First, only cascade events were kept. The track-like events are a class of background sources. The track-like events initiated by muons due to atmospheric muons and muon neutrinos would be rejected after that event selection. To reduce more background events initiated by atmospheric muon, Second, only up-going events occurring during a period in which the Sun was below the horizon were kept. Besides, only those up-going events from the Sun's direction were kept. Due to the sizable energy and angular uncertainties caused by the event reconstruction with IceCube, the cut windows for energy and angular separation between cascades and the Sun's direction would be used to extract signal candidate events from the up-going cascades events. These windows were taken to be one standard uncertainty and one median uncertainty, respectively. Certainly, the residual signals still contain a small neutrino component after all those event selections. Since the LDM and neutrino cascades are hard to distinguish at IceCube, one could only evaluate the number of expected neutrinos fallen into those windows.
\par
The factors ($C_1$ and $C_2$) should be considered in the evaluation of the numbers of expected LDMs. $C_1$ is equal to 68.3\% (that is 68.3\% of the LDM events reconstructed with IceCube fall into a window caused by one standard energy uncertainty). $C_2$ is equal to 50\% (that is 50\% of the LDM events reconstructed with IceCube fall into a window caused by one median angular uncertainty). Then the number N$_{det}$ of expected LDMs obeys the following equation:
\begin{center}
\begin{equation}
\frac{dN_{det}}{dE} =C_1\times C_2\times\int_T A_{eff}(E)\Phi_{LDM} P(E,\epsilon(t)) dt
\end{equation}
\end{center}
where $A_{eff}(E)$ obtained from the figure 2 in Ref.\cite{icecube2014a} is denoting the effective observational area for IceCube. E is denoting the energy of an incoming particle. $P(E,\epsilon(t))$ can be given by the following equation:
\begin{center}
\begin{equation}
P(E,\epsilon(t))=exp(-\displaystyle\frac{D_e(\epsilon(t))}{L_{earth}})\left(1-exp(-\displaystyle\frac{D}{L_{ice}})\right).
\end{equation}
\end{center}
where $L_{earth,ice}$ is denoting the LDM interaction lengths with the Earth and ice, respectively. D is denoting the effective length in the IceCube detector and taken to be 1 km in this work. $D_e(\epsilon(t))=2R_esin(\epsilon(t))$ is denoting the distance through the Earth. R$_e$ is denoting the radius of the Earth. $\epsilon(t)$ is denoting the obliquity of the ecliptic changing with time. The maximum value of $\epsilon$ is 23.44$^{\circ}$.
\par
After rejecting track-like events, the background remains two sources: astrophysical and atmospheric neutrinos which pass through the detector of IceCube. Only a neural current interaction with nuclei is relevant to muon neutrinos considered here. The astrophysical neutrinos flux can be described by\cite{icecube2021a}
\begin{center}
\begin{equation}
\Phi_{\nu}^{astro}=\Phi_{astro}\times\left(\displaystyle\frac{E_{\nu}}{100TeV}\right)^{-(\alpha+\beta log_{10}(\frac{E_{\nu}}{100TeV}))}\times10^{-18}GeV^{-1} cm^{-2}s^{-1}sr^{-1}
\end{equation}
\end{center}
where $\Phi_{\nu}^{astro}$ is denoting the total astrophysical neutrino flux. The coefficients, $\Phi_{astro}$, $\alpha$ and $\beta$ are given in Fig. VI.10 in Ref.\cite{icecube2021a}. The atmospheric neutrinos flux can be described by\cite{SMS}
\begin{center}
\begin{equation}
\Phi_{\nu}^{atm} = C_{\nu}\left(\displaystyle\frac{E_{\nu}}{1GeV}\right)^{-(\gamma_0+\gamma_1x+\gamma_2x^2)}GeV^{-1} cm^{-2}s^{-1}sr^{-1}
\end{equation}
\end{center}
where $x=log_{10}(E_{\nu}/1GeV)$. $\Phi_{\nu}^{atm}$ is denoting the atmospheric neutrino flux. The coefficients, $C_{\nu}$ ($\gamma_0$, $\gamma_1$ and $\gamma_2$) are given in Table III in Ref.\cite{SMS}.
\par
The neutrinos fallen into the energy and angular windows mentioned above would also be regarded as signal candidate events, so the evaluation of the number of expected neutrinos has to be performed by integrating over the region caused by these windows. Then the number of expected neutrinos N$_{\nu}$ obeys the following equation:
\begin{center}
\begin{equation}
\frac{dN_{\nu}}{dE} =\int_T \int_{\theta_{min}}^{\theta_{max}} A_{eff}(E)(\Phi_{\nu}^{astro}+\Phi_{\nu}^{atm}) P(E,\epsilon(t),\theta)\frac{2\pi r_e(\epsilon(t))^2 sin2\theta}{D_e'(\epsilon(t),\theta)^2} d\theta dt
\end{equation}
\end{center}
where $r_e(\epsilon(t))=\displaystyle\frac{D_e(\epsilon(t))}{2}$. $\theta$ is denoting the angular separation between the neutrinos and the Sun's diretion. $\theta_{min}$ = 0 and $\theta_{max}$ = $\sigma_{\theta}$. $\sigma_{\theta}$ is denoting the median angular uncertainty for cascades at IceCube. The standard energy and median angular uncertainties can be obtained from Ref.\cite{icecube2014} and Ref.\cite{icecube2017}, respectively. $P(E,\epsilon(t),\theta)$ can be given by
\begin{center}
\begin{equation}
P(E,\epsilon(t),\theta)=exp(-\displaystyle\frac{D_e'(\epsilon(t),\theta)}{L_{earth}})\left(1-exp(-\displaystyle\frac{D}{L_{ice}})\right)
\end{equation}
\end{center}
where $D_e'(\epsilon(t),\theta)=D_e(\epsilon(t))cos(\theta)$ is denoting the distance through the Earth.
\section{Results}
The distributions and numbers of expected LDMs and neutrinos were evaluated in the energy range 1-100 TeV assuming 6 years of IceCube data. Fig. 1 shows the distributions (with an energy bin of 100 GeV) of expected LDMs and neutrinos. Compared to LDMs with $m_{Z^{\prime}}$=100 GeV and $\tau_{\phi} = 5\times10^{22}$ s, the numbers of neutrino events per energy bin are at least smaller by 4 orders of magnitude in the energy range 1-10 TeV. As shown in Fig. 1, the dominant background is caused by atmospheric neutrinos in the energy range 1-5 TeV but astrophysical neutrinos at energies above about 10 TeV in this measurement.
\par
The numbers of expected neutrinos (see black solid line) are shown in Fig. 2 and 3. The evaluation of the numbers of expected neutrinos was performed by integrating over the region caused by the energy and angular windows described above. The black dash line denotes the number of expected atmospheric neutrinos. These two figures both indicate the neutrino background can be ignored in this measurement. The numbers of expected LDMs with $m_{Z^{\prime}}$=100 GeV and $\tau_{\phi} = 10^{21}$ s can reach about 70 and 1 at 1 TeV and 5.3 TeV at IceCube, respectively, as shown in Fig. 2 (see the red dash line). Fig. 2 also presents LDMs with $\tau_{\phi}=10^{22}$ s (see the blue dot line) and $\tau_{\phi}=5\times10^{22}$ s (see the magenta dash dot line) could be detected below about 3 TeV and 1.3 TeV at IceCube, respectively, when $m_{Z^{\prime}}$ = 100 GeV. Fig. 3 presents LDMs with $m_{Z^{\prime}}$=250 GeV (see the blue dot line) and $m_{Z^{\prime}}$=350 GeV (see the magenta dash dot line) could be detected below about 36 TeV and 4 TeV at IceCube, respectively, when $\tau_{\phi}=10^{21}$ s.
\section{Discussion and Conclusion}
The Ref.\cite{icecube2021b} presents an analysis of neutrino signals due to the DM annihilation in the Sun with 6 years of IceCube data. This analysis has not found any significant indication of neutrinos due to the DM annihilation in the Sun. Since the LDM and neutrino signals are hard to distinguish at IceCube, it is a reasonable assumption that no events are observed in the measurement of LDMs due to the decay of HDM in the Sun at IceCube in 6 years. The corresponding upper limit on LDM flux at 90\% C.L. was calculated with the Feldman-Cousins approach\cite{FC} (see the black solid line in Fig. 4 and 5).
\par
Fig. 4 presents the fluxes of expected LDM with $\tau_{\phi}$=10$^{21}$ s (red dash line), 10$^{22}$ s (blue dot line) and 5$\times$10$^{22}$ s (magenta dash dot line). That limit excludes the LDM fluxes with $\tau_{\phi}$ = $10^{21}$ s and $10^{22}$ s below about 4.4 TeV and 2.1 TeV, respectively. Then the LDMs could be probed with $\tau_{\phi}\lesssim 5\times10^{22}$ s at IceCube.
\par
Fig. 5 shows the fluxes of expected LDM with $m_{Z^{\prime}}$=100 GeV (red dash line), 250 GeV (blue dot line) and 350 GeV (magenta dash dot line). That limit excludes the LDM fluxes with $m_{Z^{\prime}}$ = 100 GeV and 250 GeV below about 4.4 TeV and 10.3 TeV, respectively. Then the LDMs could be probed with 100 GeV $\lesssim m_{Z^{\prime}}\lesssim$ 350 GeV at IceCube.
\par
Based on the results described above, it is a reasonable conclusion that those LDMs could be directly detected in the O(1TeV)-O(10TeV) energy range at IceCube with 100 GeV $\lesssim m_{Z^{\prime}} \lesssim$ 350 GeV and $\tau_{\phi} \lesssim 5\times10^{22}$ s. Since these constraints are only given by the assumptions mentioned above, certainly, the experimental collaborations, like the IceCube collaboration, should be encouraged to conduct an unbiased analysis with the data of IceCube.
\par
Since $\Phi_{LDM}$ is proportional to $\displaystyle\frac{1}{\tau_{\phi}}$ (see Eqn. 10), the above results actually depends on the lifetime of heavy DM, $\tau_{\phi}$. If $\tau_{\phi}$ varies from 10$^{18}$ s to 10$^{20}$ s, the numbers of expected LDMs with IceCube are larger by from 3 to 1 orders of magnitude than that with $\tau_{\phi}=10^{21}$ s, respectively.
\par
Besides, the capability of the measurement of those LDMs was roughly evaluated with the ANTARES telescope. Those LDMs could be directly detected at energies with O(1TeV) at ANTARES with $m_{Z^{\prime}}$ < 200 GeV and $\tau_{\phi} < 10^{20}$ s. Compared to IceCube, the expected signal to background rate is larger by about one order of magnitude with ANTARES. It is more difficult for ANTARES to detect those LDMs, however, since the effective area for ANTARES is smaller by about 2 orders of magnitude than that for IceCube at energies above 1 TeV\cite{icecube2014a,antares-icecube2015}. As we all know, the capabilities of the measurement of those LDMs should be substantially improved with IceCube and ANTARES if their upgrading projects will be completed in the future.
\par
Ref.\cite{APQ} presents an analysis of the constraints on the mass of $Z^{\prime}$ and $g_{\chi\chi Z^{\prime}}$ using the observations of direct, indirect measurement, collider and cosmology. Considered the assumption that the bulk of present-day DM consists of HDMs in this work, observations of cosmology and direct, indirect measurements of DM are inappropriate to analysis the results in this work. Fig. 6 in Ref.\cite{APQ} shows the result for DM with a mass of 8 GeV. The whole light $Z^{\prime}$ window (m$_{Z^{\prime}}$ < 1 TeV) is ruled out by the observations of LHC and Tevatron using the dijet data. To measure DM generated by colliders, the dijet+E$^{miss}_T$ analysis is more reasonable. Fig. 6 in Ref.\cite{APQ} presents the light $Z^{\prime}$ window is not ruled out by LHC at 8 TeV using the dijet+E$^{miss}_T$ data with $g_{\chi\chi Z^{\prime}}$ < 0.25.
\section{Acknowledgements}
This work was supported by the National Natural Science Foundation
of China (NSFC) under the contract No. 11235006, the Science Fund of
Fujian University of Technology under the contracts No. GY-Z14061 and GY-Z13114 and the Natural Science Foundation of
Fujian Province in China under the contract No. 2015J01577.
\par

\newpage

\begin{figure}
 \centering
 \includegraphics[width=0.9\textwidth]{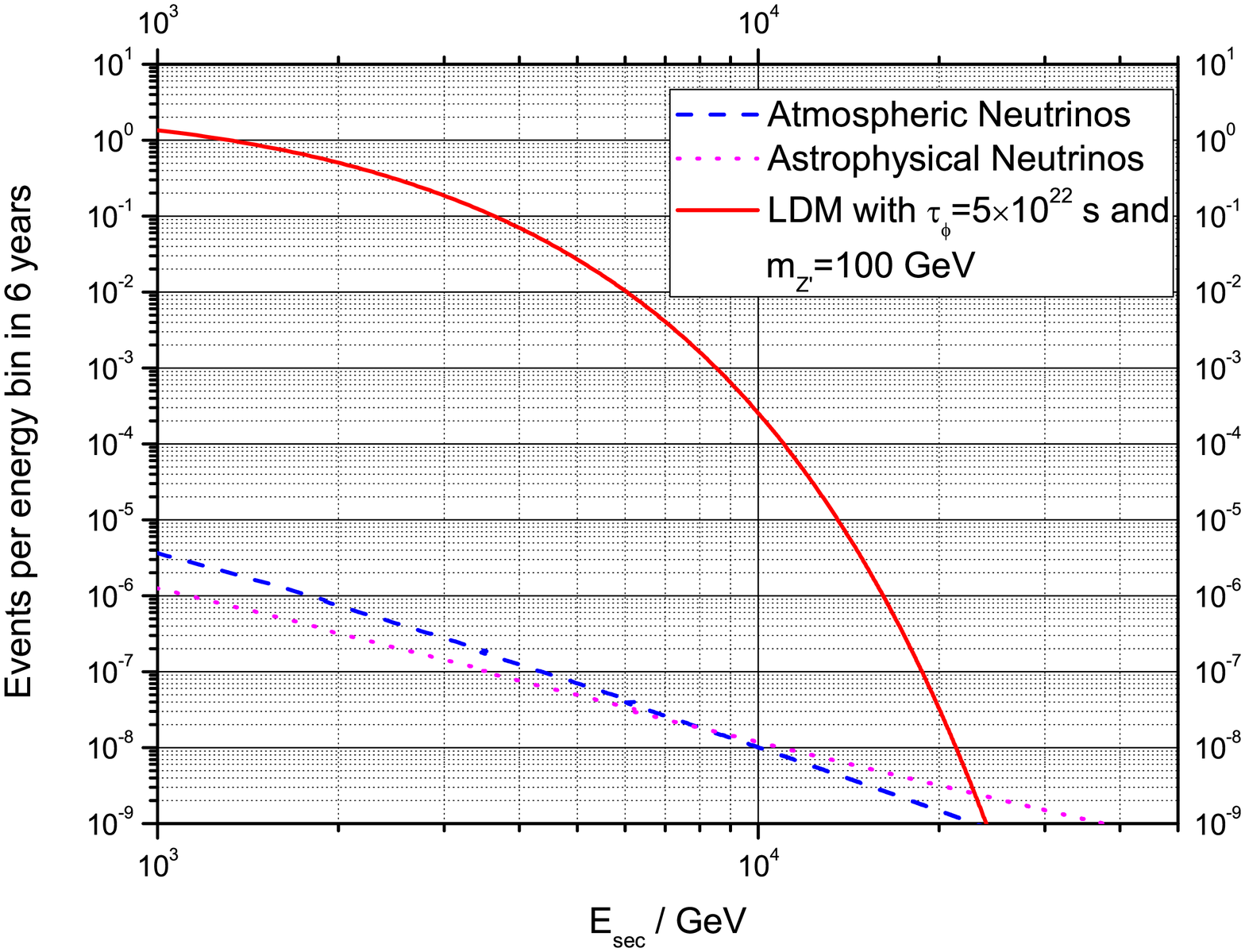}
%%% bb = left_bottom_X, left_bottom_Y, right_top_X, right_top_Y
%%% scale through "set width"
 \caption{Distributions of expected LDMs with $\tau_{\phi}$ = $5\times10^{22}$ s and $m_{Z^{\prime}}$ = 100 GeV and neutrinos. Their energy bins are 100 GeV.}
 \label{fig:distribution}
\end{figure}

\begin{figure}
 \centering
 %\includegraphics[bb=0 0 200 300, width=3.8cm]{energy}
 %\hspace{0.7\textwidth}
 \includegraphics[width=0.9\textwidth]{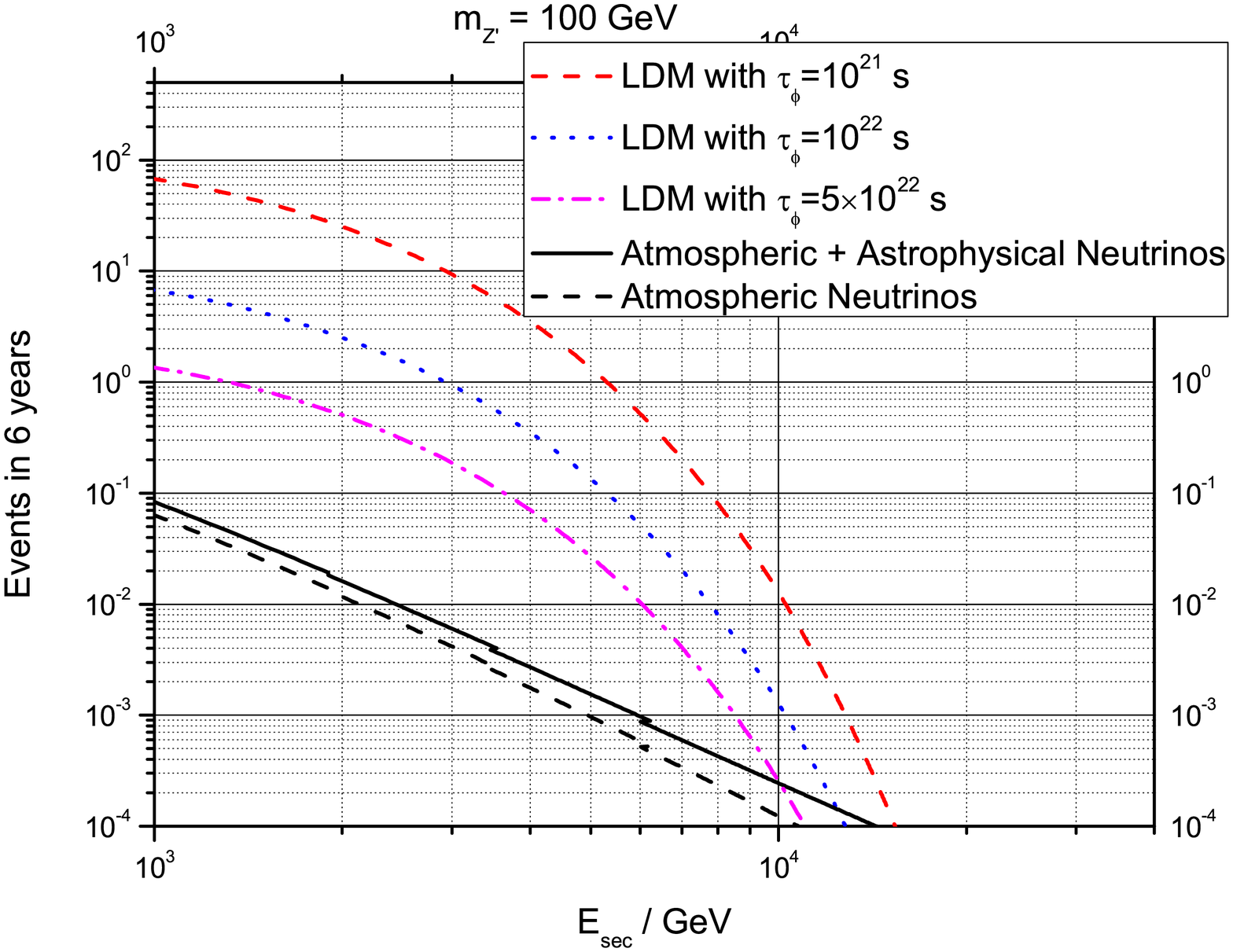}
 \caption{With the different $\tau_{\phi}$ ($10^{21}$ s, $10^{22}$ s and $5\times10^{22}$ s), the numbers of expected LDMs were evaluated assuming 6 years of IceCube data, respectively. The evaluation of numbers of expected neutrinos was performed by integrating over the regions caused by one standard energy and median angular uncertainties.}
 \label{fig:tau_100GeV}
\end{figure}

\begin{figure}
 \centering
 \includegraphics[width=0.9\textwidth]{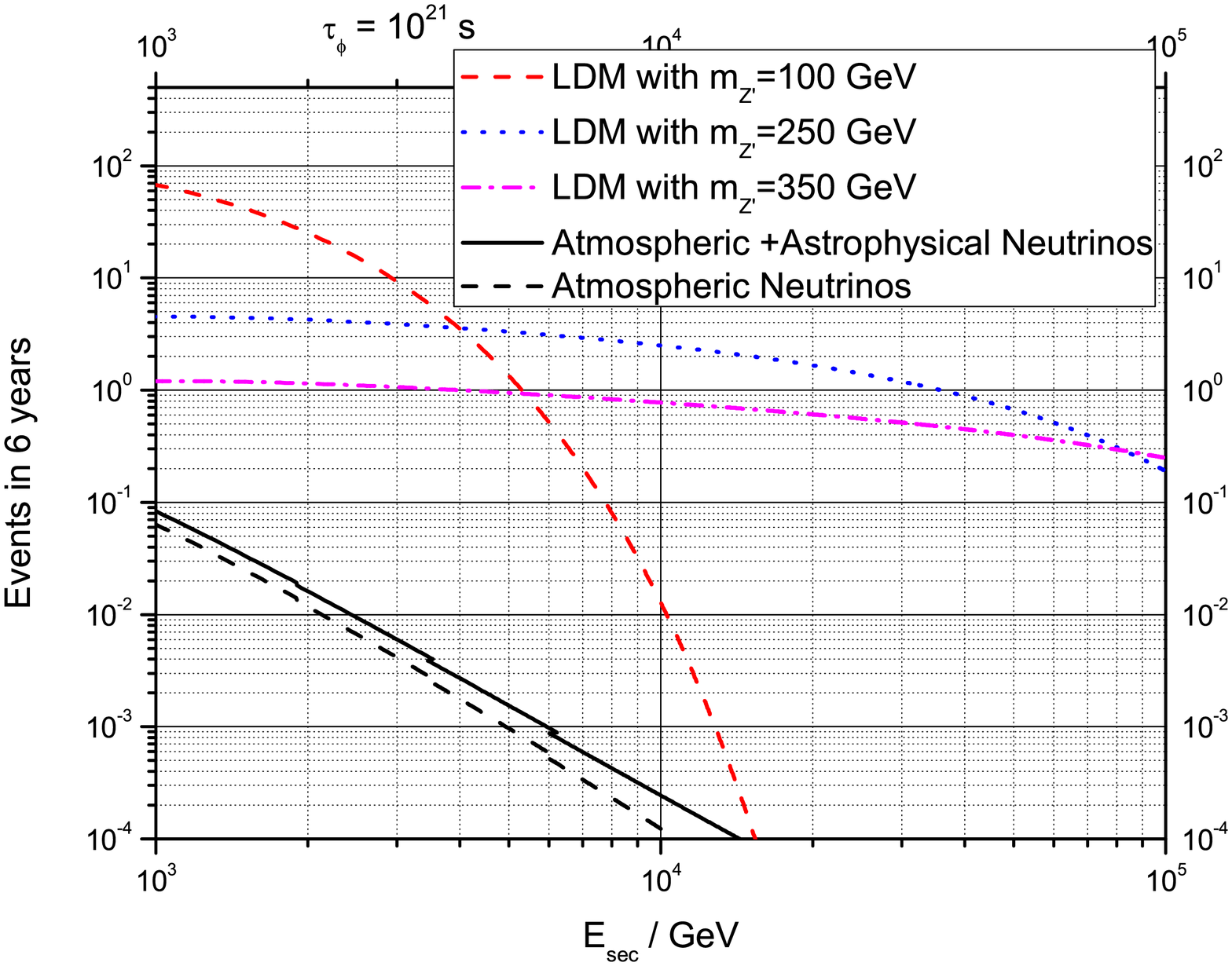}
%%% bb = left_bottom_X, left_bottom_Y, right_top_X, right_top_Y
%%% scale through "set width"
 \caption{With the different $Z^{\prime}$ masses (100 GeV, 250 GeV and 350 GeV), the numbers of expected LDMs were evaluated assuming 6 years of IceCube data, respectively. The evaluation of numbers of expected neutrinos was performed by integrating over the regions caused by one standard energy and median angular uncertainties.}
 \label{fig:mz_1e21}
\end{figure}

\begin{figure}
 \centering
 %\includegraphics[bb=0 0 200 300, width=3.8cm]{energy}
 %\hspace{0.7\textwidth}
 \includegraphics[width=0.9\textwidth]{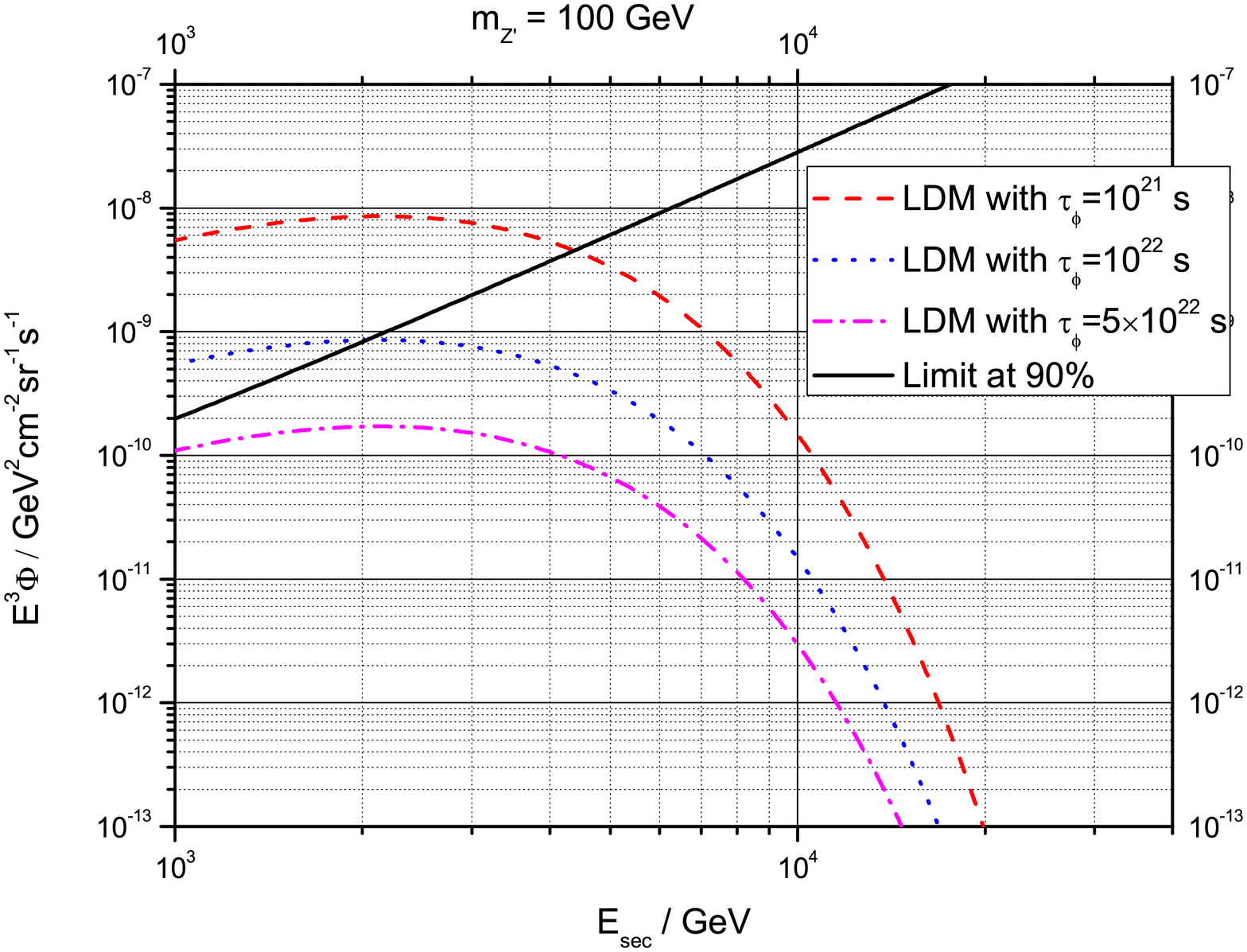}
 \caption{With the different $\tau_{\phi}$ ($10^{21}$ s, $10^{22}$ s and $5\times10^{22}$ s), the fluxes of expected LDM were estimated at IceCube, respectively. Assuming no observation at IceCube in 6 years, the upper limit at 90\% C.L. was also computed.}
 \label{fig:flux_tau_100GeV}
\end{figure}

\begin{figure}
 \centering
 %\includegraphics[bb=0 0 200 300, width=3.8cm]{energy}
 %\hspace{0.7\textwidth}
 \includegraphics[width=0.9\textwidth]{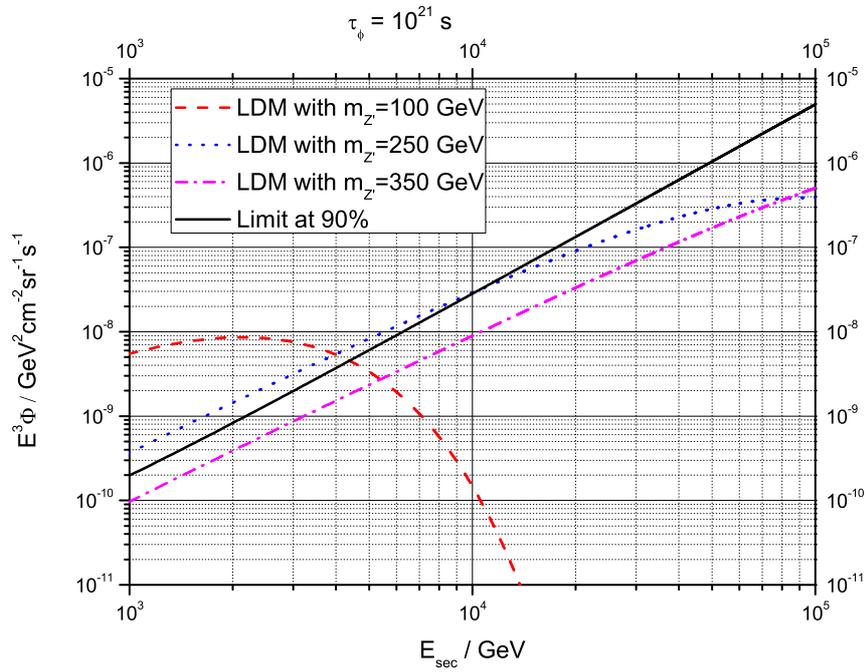}
 \caption{With the different $Z^{\prime}$ masses (100 GeV, 250 GeV and 350 GeV), the fluxes of expected LDM were estimated at IceCube, respectively. Assuming no observation at IceCube in 6 years, the upper limit at 90\% C.L. was also computed.}
 \label{fig:flux_mz_1e21}
\end{figure}

\end{document}